\documentclass[twocolumn,a4paper,pra,aps,10pt,superscriptaddress]{revtex4-2}

\usepackage[utf8x]{inputenc} 
\usepackage{amsfonts}
\usepackage{amsmath}
\usepackage{graphicx} 
\usepackage{dcolumn} 
\usepackage{bm} 
\usepackage{braket} 
\usepackage[hidelinks]{hyperref}
\usepackage[per-mode=symbol]{siunitx} 
\usepackage[capitalize]{cleveref} 

\let\originalleft\left
\let\originalright\right
\renewcommand{\left}{\mathopen{}\mathclose\bgroup\originalleft}
\renewcommand{\right}{\aftergroup\egroup\originalright}

\let\oldcite\cite
\renewcommand{\cite}[1]{\mbox{\oldcite{#1}}}

\begin{document}

\date{\today}

\title{Intermodulation Distortion in a Josephson Traveling Wave Parametric Amplifier}

\author{Ants Remm}
\affiliation{Department of Physics, ETH Zurich, CH-8093 Zurich, Switzerland}
\author{Sebastian Krinner}
\affiliation{Department of Physics, ETH Zurich, CH-8093 Zurich, Switzerland}
\author{Nathan Lacroix}
\affiliation{Department of Physics, ETH Zurich, CH-8093 Zurich, Switzerland}
\author{Christoph Hellings}
\affiliation{Department of Physics, ETH Zurich, CH-8093 Zurich, Switzerland}
\author{Francois Swiadek}
\affiliation{Department of Physics, ETH Zurich, CH-8093 Zurich, Switzerland}
\author{Graham Norris}
\affiliation{Department of Physics, ETH Zurich, CH-8093 Zurich, Switzerland}
\author{Christopher Eichler}
\affiliation{Department of Physics, ETH Zurich, CH-8093 Zurich, Switzerland}
\author{Andreas Wallraff}
\affiliation{Department of Physics, ETH Zurich, CH-8093 Zurich, Switzerland}
\affiliation{Quantum Center, ETH Zurich, CH-8093 Zurich, Switzerland}

\begin{abstract}
	Josephson traveling wave parametric amplifiers enable the amplification of weak microwave signals close to the quantum limit with large bandwidth, which has a broad range of applications in superconducting quantum computing and in the operation of single-photon detectors.
	While the large bandwidth allows for their use in frequency-multiplexed detection architectures, an increased number of readout tones per amplifier puts more stringent requirements on the dynamic range to avoid saturation.
	Here, we characterize the undesired mixing processes between the different frequency-multiplexed tones applied to a Josephson traveling wave parametric amplifier, a phenomenon also known as intermodulation distortion.
	The effect becomes particularly significant when the amplifier is operated close to its saturation power.
	Furthermore, we demonstrate that intermodulation distortion can lead to significant crosstalk and reduction of fidelity for multiplexed readout of superconducting qubits.
	We suggest using large detunings between the pump and signal frequencies to mitigate crosstalk.
	Our work provides insights into the limitations of current Josephson traveling wave parametric amplifiers and highlights the importance of performing further research on these devices.
\end{abstract}

\maketitle

\section{Introduction} \label{sec:introduction}

Amplification of weak microwave signals is essential for many applications, including readout of superconducting qubits~\cite{Blais2021,Kjaergaard2020a,Wallraff2005,Mallet2009} and quantum dot devices~\cite{Petersson2012,Zheng2019,deJong2021}, and radio astronomy~\cite{Day2003,Zobrist2019}. 
State-of-the-art low-noise amplifiers in the microwave domain~\cite{Yurke1996,Castellanos2007, Castellanos2008, Eichler2014a, Macklin2015, HoEom2012} approach the quantum limit in noise performance~\cite{Caves1982} by operating at millikelvin temperatures and using parametric pumping of a nonlinear circuit made of Josephson junctions or high kinetic inductance elements.
While parametric amplifiers based on nonlinear resonators have typical bandwidths on the order of tens of MHz limited by the gain and the resonator linewidth~\cite{Eichler2014,Planat2019}, traveling wave parametric amplifiers~(TWPAs)~\cite{Yurke1996,O'Brien2014a,Macklin2015,White2015,Planat2020,Esposito2021,Ranadive2022} can have much higher bandwidths of up to several GHz.
The high bandwidth enables a high degree of frequency-multiplexing, for example for qubit readout~\cite{Heinsoo2018,deJong2021} and single-photon detectors~\cite{Zobrist2019}.
Multiplexed use of hardware resources~\cite{Chen2012f,Jerger2012,Barends2014,Schmitt2014a,Heinsoo2018,Arute2019,Krinner2022} is essential for the operation of large quantum devices.

So far material losses and the generation of signal sidebands have been identified as the main sources of excess noise in TWPAs above the quantum limit~\cite{Macklin2015,Esposito2021,Peng2022}, characterized by the intrinsic quantum efficiency.
In addition to adding as little noise as possible at the signal frequency, broadband amplifiers should not generate spurious tones due to intermodulation of the inputs~\cite{Frattini2018b,Sivak2019}, in particular when they are used in frequency-multiplexed applications.
However, due to amplifier nonlinearities, intermodulation distortion is unavoidable and constitutes a well-known phenomenon in classical amplifiers~\cite{Walker2011}.
Intermodulation products can lead to crosstalk between the amplified signals if any of their frequencies overlap with one of the signals.
The probability of such collisions increases with increasing degree of frequency multiplexing because the number of intermodulation products increases.

In this work we characterize the intermodulation distortion of a resonantly phase-matched traveling wave parametric amplifier~\cite{Macklin2015} (\cref{sec:characterization}). 
We identify intermodulation products of order up to five in the output spectrum and characterize their power and frequency dependence on the input signals.
We then show that the frequency collision of an intermodulation product with a readout signal can lead to significant crosstalk and reduction of readout fidelity (\cref{sec:implications}). 
Finally, we discuss strategies to mitigate these errors by the choice of pump and signal frequency and of power levels (\cref{sec:mitigation}).

We find that intermodulation distortion, if not accounted for, can lead to significant crosstalk and readout errors in frequency-multiplexed architectures, which can be detrimental for applications that rely on fast, high-fidelity readout, such as quantum error correction~\cite{Dennis2002,Kitaev2003,Raussendorf2007,Fowler2012}. 
Therefore, continued development of cryogenic microwave amplifiers is needed to allow for fast low-crosstalk multiplexed readout of many qubits.

\section{Intermodulation Distortion Characterization} \label{sec:characterization}

\begin{figure}
	\includegraphics[width=\columnwidth]{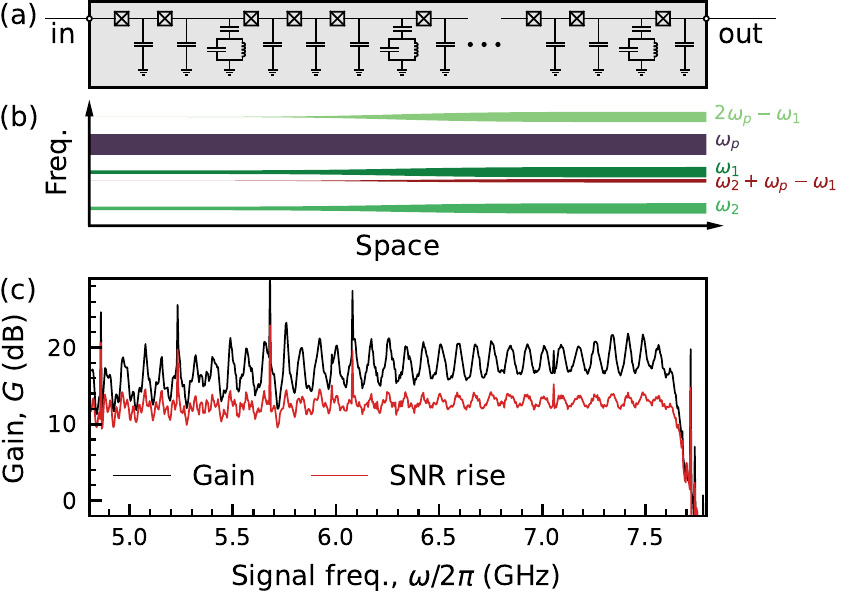}
	\caption{(a)~The circuit of a resonantly phase-matched TWPA. (b)~Examples of tones propagating along the length of the TWPA and their amplitudes (thickness of the lines, not to scale) in the case of frequency-multiplexed readout at $\omega_1$ and $\omega_2$ with the pump tone at $\omega_p$. In particular, a tone at $\omega_2+\omega_p-\omega_1$ is created, which is an example of the intermodulation products studied in this work. (c)~The measured gain $G$ (black) and signal-to-noise ratio improvement (red) spectra of the device under study.}
	\label{fig:concept}
\end{figure}

The TWPA investigated in this study is a nonlinear transmission line, formed by a series of Josephson junctions and capacitors to ground, which is shunted to ground by capacitively coupled LC resonators at \SI{8.1}{GHz} to achieve phase matching~\cite{Macklin2015}, see \cref{fig:concept}~(a).
The nonlinearity of the Josephson junctions allows for four-and-higher-wave mixing processes between the tones traversing the TWPA.
In the presence of a strong pump tone, the mixing processes lead to desired phase-insensitive amplification of weak tones [$\omega_1$, $\omega_2$ in \cref{fig:concept}~(b)], but also to sideband generation by absorbing or emitting pump photons~\cite{Peng2022} and to the mixing of multiple signals~\cite{Frattini2018b}, known as intermodulation distortion. 
An example of intermodulation distortion is the creation of a tone (intermodulation product) at frequency $\omega_2 + \omega_p - \omega_1$ at the output of the TWPA, as shown in~\cref{fig:concept}~(b).

In our experimental setup the TWPA is mounted at the base temperature stage of a dilution cryostat operated at \SI{10}{mK}. 
The signals, generated at room temperature, are up-converted to microwave frequencies using an IQ-mixer.
The output of the TWPA is amplified by a cryogenic high-electron mobility transistor (HEMT) amplifier and by room-temperature amplifiers. 
Finally, the signals are down-converted using an IQ-mixer with a local oscillator at \SI{6.9}{GHz} and digitized, see \cref{supp:setup} for details.
The power \SI{-62}{dBm} and frequency $\omega_p/2\pi = \SI{7.92}{GHz}$ of the pump tone are chosen to maximize the signal-to-noise ratio of signals applied between \SI{6.7}{GHz} and \SI{7.6}{GHz} used for qubit readout.
The power levels at the input of the TWPA are calculated based on room-temperature measurements and the attenuation of the components within the cryostat.
We achieve a mean gain of $G=\SI{18.4}{dB}$ and a signal-to-noise ratio rise of \SI{13.0}{dB} relative to not pumping the TWPA, in which case the noise is dominated by the HEMT amplifier, see \cref{fig:concept}~(c).

\begin{figure}
	\includegraphics[width=\columnwidth]{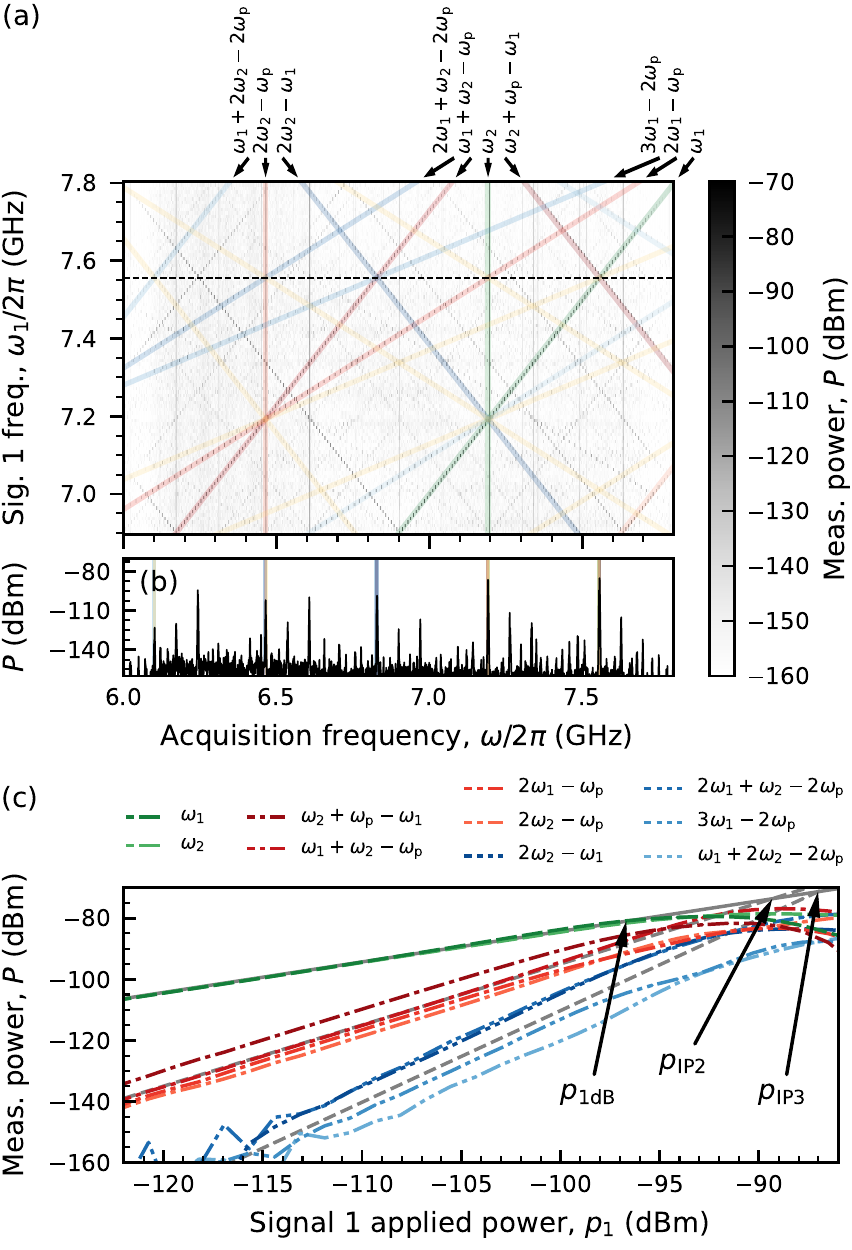}
	\caption{(a)~Measured output power spectra when sweeping the frequency of a single tone $\omega_1$ in the presence of a second signal tone $\omega_2$. Various intermodulation products are highlighted in color according to their signal order, see text for details. (b)~Line cut of the data in (a) at $\omega_1$ indicated by the dashed line. (c)~Dependence of the power of the intermodulation products $P$ on the applied signal power $p_1$. The power of the two signal tones is swept together. The gray lines indicate the mean power level for products of signal order $O_s=1$, 2, and 3 according to \cref{eq:intermodulation-power-model}, and the arrows indicate the \SI{1}{dB} compression power $p_\mathrm{1dB}$ and the intermodulation intercept points $p_\mathrm{IP}$ for $O_s=2$ and 3 tones.}
	\label{fig:characterization}
\end{figure}

To observe the intermodulation products that we want to characterize, we apply signal tones at frequencies $\omega_1$ and $\omega_2$ to the input of the TWPA through a directional coupler which is also used for applying the pump tone.
The signal power level at the input of the TWPA is approximately \SI{-102}{dBm}.
We acquire \SI{2.275}{\micro s} long time traces and multiply them with a Blackman-Harris window~\cite{Harris1978} before taking the Fourier transform to avoid windowing effects.
By fixing $\omega_2/2\pi = \SI{7.1924}{GHz}$ and sweeping $\omega_1/2\pi$ from \SIrange{6.9}{7.8}{GHz}, we record the spectra shown in \cref{fig:characterization}~(a).
In addition to the signals at $\omega_1$ and $\omega_2$ (green lines), we observe intermodulation products at frequencies $\omega = n_p \omega_p + \sum_{i=1,2} n_i \omega_i$ with integers $n_p$ and $n_i$.
We can classify the intermodulation products according to their total order $O_t = |n_p| + \sum_i |n_i|$.
Four-wave mixing processes lead to products of total order $O_t = 3$ while cascaded four-wave mixing processes and higher-order mixing processes can lead to products with odd order $O_t = 5$ and higher.
Of the intermodulation products that lie within the acquisition band, we observe all 17 products with total order $O_t = 3$ or 5 and several products with $O_t = 7$ above the noise floor of \SI{-160}{dBm} (at the output of the TWPA).
Potential intermodulation products of even total order fall outside the acquisition bandwidth in our setup.
The presence of all the intermodulation products implies that there are no selection rules determining which intermodulation products can be created, other than the parity of~$O_t$.
While the allowed total orders are determined by the order of the mixing process, the power level of the intermodulation product is mostly determined by the signal order $O_s = \sum_i |n_i|$, as we will see below.
The intermodulation products of $O_s = 1$, 2, 3, and $\ge4$ are highlighted in \cref{fig:characterization}~(a) in green, red, blue, and yellow, respectively.
Each intermodulation product also appears mirrored around the local oscillator frequency due to imperfections of the frequency down-conversion process.

Next, we investigate the power of the intermodulation products $P$.
We fix the frequency $\omega_1/2\pi = \SI{7.5551}{GHz}$, as indicated by the linecut in \cref{fig:characterization}~(b), and vary the input powers $p_1$ and $p_2 = 0.5 p_1$ of the two signals, chosen such that the output powers are similar.
We record the power $P$ of the intermodulation products of $O_s \le 3$, see \cref{fig:characterization}~(c), and find that they follow power laws with the signal order $O_s$ as the exponent as long as the amplifier is not saturated.
Comparing the powers of different $O_s=3$ products, we find that they can be of similar magnitude even for different $O_t$ values.
The observed power-laws together with the data from independent power sweeps of the two tones (see \cref{supp:separate-powersweep}) motivate an empirical model of the output power
\begin{equation} \label{eq:intermodulation-power-model}
	P = G p_{\mathrm{IP}} \prod_i \left(\frac{p_i}{p_{\mathrm{IP}}}\right)^{|n_i|}.
\end{equation}
The model is parametrized by the mean gain $G$ and the intermodulation distortion intercept point $p_\mathrm{IP}$~\cite{Walker2011}, i.e., the input power level at which the extrapolated intermodulation product power equals the signal power.
Using the average gain of the two signals $G=\SI{17.2\pm1.3}{dB}$, we calculate $p_\mathrm{IP}$ for each intermodulation product according to \cref{eq:intermodulation-power-model} at input power $p_1 = \SI{-106}{dBm}$, which is significantly below the saturation power.
We find a mean second order intercept point $p_\mathrm{IP2} = \SI{-91\pm3}{dBm}$ (for $O_s = 2$) and a mean third order intercept point $p_\mathrm{IP3} = \SI{-88\pm3}{dBm}$ (for $O_s = 3$), see dashed gray lines in \cref{fig:characterization}~(c) calculated according to \cref{eq:intermodulation-power-model} and the mean $p_\mathrm{IP}$ values.
The uncertainties indicate one standard deviation of the spread over different intermodulation products.
The $p_\mathrm{IP}$ values, visualized as the intercepts of the gray dashed lines with the solid gray line (mean signal power) in \cref{fig:characterization}~(c), are close to the \SI{1}{dB} compression power $p_\mathrm{1dB} = \SI{-96.7\pm2.3}{dBm}$ (see \cref{supp:saturation}).
The power differences between intermodulation products of the same signal order might be due to differences in the conversion rates or due to the frequency-dependence of the gain.

We can use a simple model to describe the relation between the \SI{1}{dB} compression power $p_\mathrm{1dB}$ and the third order intermodulation intercept power $p_\mathrm{IP3}$. 
In the lowest-order series expansion that can explain four-wave mixing, we write the output voltage of the amplifier as $V_\mathrm{out} = \sqrt{G} V_\mathrm{in}\left(1 - k V_\mathrm{in}^2\right)$, for input voltage $V_\mathrm{in}$ and a coefficient $k$ which determines both saturation and intermodulation properties of the amplifier.
From this model, we find $p_\mathrm{IP3}/p_\mathrm{1dB} = \SI{9.6}{dB}$~\cite{Walker2011}, similar to the observed $p_\mathrm{IP3}/p_\mathrm{1dB} = \SI{9\pm4}{dB}$. 
We therefore expect that the intermodulation distortion intercept powers increase if the \SI{1}{dB} compression power of the amplifier is increased.

\section{Implications for multiplexed readout} \label{sec:implications}

We assess the impact of intermodulation distortion on the performance of frequency-multiplexed qutrit readout using the device presented in Krinner \textit{et al.}~\cite{Krinner2022}.
Specifically, we study how frequency multiplexing affects the signal-to-noise ratio, and how intermodulation products can lead to crosstalk and increased readout errors.

\begin{figure}
	\includegraphics[width=\columnwidth]{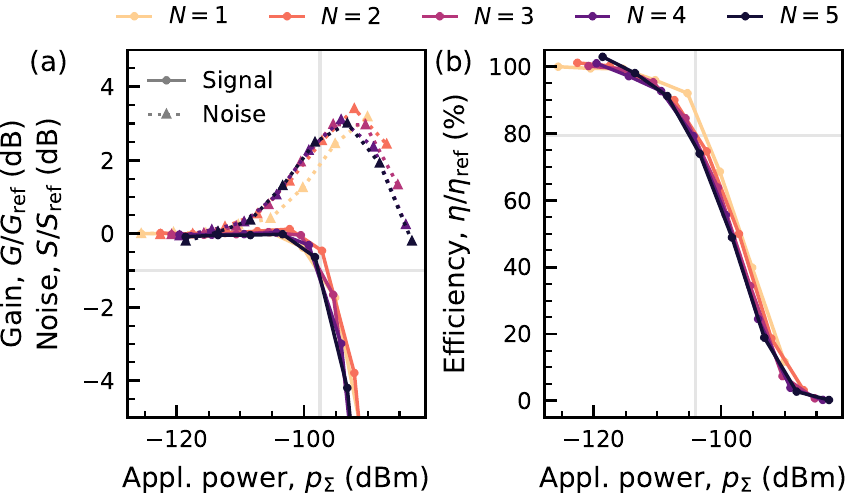}
	\caption{(a)~Change in the signal gain $G$ and noise power $S$ as a function of applied power $p_\Sigma$ and number of signal tones $N$, compared to the single-signal ($N=1$), low-power ($p_\Sigma=\SI{-126}{dBm}$) values. (b)~Change in measurement efficiency $\eta$ relative to the single-signal, low-power value $\eta_\mathrm{ref}$. The intersection points of the gray lines indicate the \SI{1}{dB} gain loss in panel (a) and \SI{1}{dB} efficiency loss in panel (b).}
	\label{fig:efficiency}
\end{figure}

To investigate the performance of the TWPA in the presence of multiple input tones, we apply 31 different subsets of five frequency components $\omega_i/2\pi = \{7.5551, 7.1924, 7.3725, 6.979, 6.76076\}_i~\si{GHz}$ that could be used for multiplexed readout of five qubits.
We scale the power of the all the applied tones by a common factor and record time traces.
We calculate the signal gain as $G_i = \left|\langle A_i \rangle\right|^2/p_i$ and noise as $S_i = \left\langle |A_i|^2 \right\rangle - \left|\langle A_i \rangle\right|^2$, where $A_i$ is the integrated amplitude of the timetrace, down-converted from frequency $\omega_i$, $p_i$ is the applied power, and the averaging is done over $2^{10}$ acquired time traces.
To find the average signal gain and noise for a given degree of multiplexing $N$, we average $G_i$ and $S_i$ over all the frequency components $i$ and subsets of $N$ frequency components that include that component $i$.
We compare the average gain and noise to the low-power single-tone reference values $G_\mathrm{ref}$ and $S_\mathrm{ref}$.

Both gain and noise rise mainly depend on the total applied power $p_\Sigma$, and only weakly on the number of tones $N$, as seen in \cref{fig:efficiency}~(a) by the collapse of all curves for different $N$ on a single one.
This means that the creation of the additional intermodulation products does not significantly reduce the gain nor increase the noise at the signal frequencies as long as the products do not overlap with the signals.
The weak dependence on the number of tones might be due to the balance of two mechanisms affecting pump depletion in opposite directions.
While the number of intermodulation products created increases with $N$, the power of each one decreases due to the lower powers of the input signals $p_\Sigma/N$.

We also analyze the saturation performance in terms of the amplifier efficiency $\eta$, i.e., the signal-to-noise ratio at its output, relative to the standard quantum limit~\cite{Caves1982}, see \cref{fig:efficiency}~(b).
From a separate characterization measurement~\cite{Bultink2018}, we find a single-signal, low-power measurement efficiency of the total detection line of $\eta_\mathrm{ref}=\num{24\pm5}\%$ (on a scale where an ideal phase-preserving amplifier would have $\eta=1$).
When operating the amplifier close to its saturation power, we observe that the noise begins to rise at input powers about \SI{10}{dB} lower than the value at which gain is significantly reduced. 
This is reflected by a \SI{1}{dB} reduction of efficiency already \SI{6.5}{dB} before the \SI{1}{dB} gain compression point is reached.
While the origin of the noise rise needs further investigation, we conclude that it is not sufficient to consider only the gain and the \SI{1}{dB} compression power, but also the noise rise when operating a TWPA close to saturation.

\begin{figure}
	\includegraphics[width=\columnwidth]{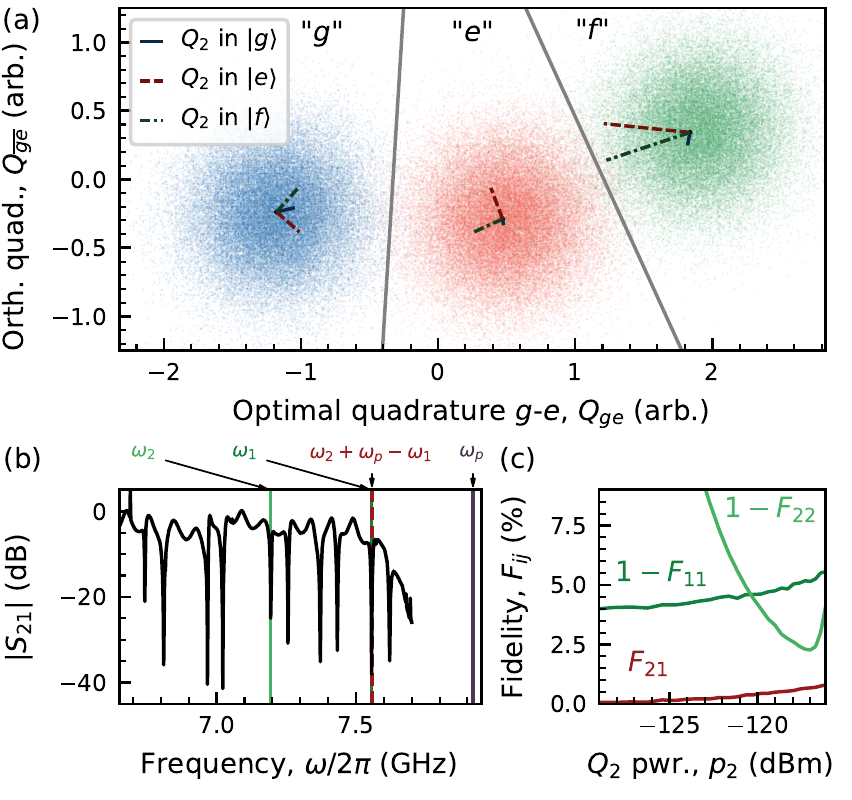}
	\caption{
		(a)~Readout histograms of Q1 when prepared in each of the three first transmon states $\ket{g}$, $\ket{e}$, and $\ket{f}$. The dash-dotted lines indicate the shift ($\times 4$ for improved visibility) of the mean of the Gaussian distribution of measurement outcomes, when Q2 is prepared in one of the qutrit states and read out in parallel. The gray solid lines indicate classification thresholds.
		(b)~Resonator spectrum of the feedline under inspection. The intermodulation product $\omega_2+\omega_p-\omega_1$ is \SI{2.2}{MHz} detuned from the readout tone of Q1 at $\omega_1$.
		(c)~Readout errors of the two qutrits, $1-F_{11}$ and $1-F_{22}$, and the readout cross-fidelity $F_{21}$, as a function of the average readout power of Q2 at the TWPA input $p_2$. Panel (a) was recorded at $p_2 = \SI{-120}{dBm}$.}
	\label{fig:readout}
\end{figure}

Beyond reducing efficiency, an intermodulation product can overlap with a signal tone used for frequency-multiplexed readout of a transmon qubit.
To investigate this effect, we use qubits X2 and D5 from the device presented in Krinner \textit{et al.}~\cite{Krinner2022}, which we label Q1 and Q2 here, respectively, see \cref{supp:qubit-device}. 
The qubit frequencies during readout are \SI{5.89}{GHz} and \SI{5.15}{GHz}, and the readout frequencies are $\omega_1/2\pi = \SI{7.5551}{GHz}$ and $\omega_2/2\pi = \SI{7.1924}{GHz}$ as before, see \cref{fig:readout}~(b) for the transmission spectrum of the readout feedline.

First, we prepare Q1 in one of its three lowest-lying energy eigenstates, labeled $\ket{g}$, $\ket{e}$, and $\ket{f}$, and then read it out using a 200-ns-long Gaussian-filtered ($\sigma = \SI{10}{ns}$) square pulse.
We integrate each acquired time trace with two sets of orthogonal mode-matched integration weights~\cite{Krinner2022}, leading to the values $Q_{ge}$ and $Q_{\overline{ge}}$. 
We use a Gaussian mixture model to classify the outcome as one of the qutrit states, see \cref{fig:readout}~(a).

We then read out Q2 simultaneously with Q1, preparing it in one of the three qutrit states, in which case the intermodulation product $\omega_2 + \omega_p - \omega_1$ is created at $\omega_1 + 2\pi \times \SI{2.2}{MHz}$.
The \SI{2.2}{MHz} detuning is small compared to the bandwidth of the \SI{200}{ns} readout pulse, but comparable to the bandwidth of the \SI{2.275}{\micro s} pulse used for characterization in \cref{sec:characterization}.
As the intermodulation product lies within the acquisition band of Q1, the centers of the Gaussian distributions are shifted depending on the state of Q2, as indicated in \cref{fig:readout}~(a).
Because the amplitude and phase of the intermodulation product depend on both the tone at $\omega_1$ and at $\omega_2$ and thereby the state of both qutrits, we observe a different shift of the mean readout response for each pair of qutrit states.
The shifts can therefore not be corrected by a linear correction operation and the crosstalk leads to a reduction of readout fidelity.

We sweep the readout power of Q2 while reading out the two qutrits in parallel, and measure the readout cross-fidelity matrix $F_{ij}$, that is, the readout fidelity of qutrit Q$i$ when looking at the classified outcome of Q$j$, normalized such that $F_{ij} = 0$ for random assignment~\cite{Heinsoo2018}, see \cref{supp:cross-fidelity}.
For perfect readout, the off-diagonal terms would be zero, while the diagonal terms would be one.
A non-zero off-diagonal term $F_{21}$ means that we can get some information about the state of Q2 from the readout of Q1, a clear indication of crosstalk.
We see in \cref{fig:readout}~(c), that the readout error of Q2, $1-F_{22}$,  decreases as we increase the readout power $p_2$, but at the same time the readout error of Q1, $1-F_{11}$, increases as does the cross-fidelity $F_{21}$.
This highlights the trade-off between high-fidelity and low-crosstalk readout.
The probability and impact of such frequency-collisions increases with the speed of the readout, as the acquisition bandwidth needs to be wider and the signal power levels higher, implying that the relative amplitude of the intermodulation products is also higher according to~\cref{eq:intermodulation-power-model}.

\section{Mitigation strategies} \label{sec:mitigation}

We can categorize the methods to reduce the crosstalk and readout errors from intermodulation in the TWPA into two broad classes.
First, one can accept that there will be frequency-collisions with the intermodulation products and try to reduce the relative amplitudes of the spurs compared to the signal.
Second, one can try to avoid frequency-collisions with the intermodulation products.

The most practical way to reduce amplitudes of intermodulation products is to increase the intercept powers $p_\mathrm{IP}$ by increasing the saturation power~\cite{Eichler2014}, as we discussed in \cref{sec:characterization}.
For example, similar TWPAs are available with a nominal \SI{1}{dB} compression power of $p_\mathrm{1dB}=-\SI{85}{dBm}$.
Alternatively, the power applied to the amplifier could be reduced.
This could be done for example by optimizing the ratio of dispersive shift and resonator linewidth (see \cref{supp:power-efficiency}), or by interferometrically canceling the mean response of the system by displacing the input field via the directional coupler that is used to add the pump tone to the TWPA input.

\begin{figure}
	\includegraphics[width=\columnwidth]{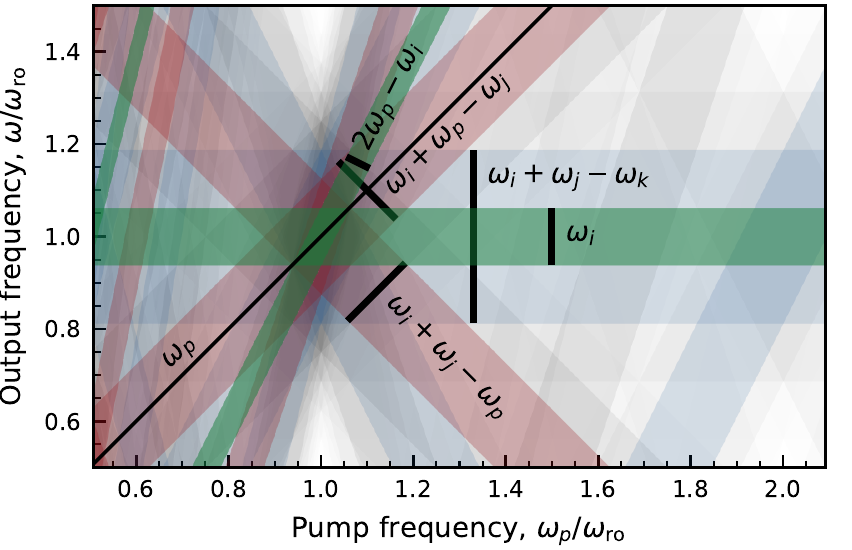}
	\caption{
		Frequencies $\omega$ of different intermodulation product classes as a function of the pump frequency $\omega_p$. The applied readout tones $\omega_{i},\omega_{j}$, and $\omega_{k}$ can have any frequency within the readout band (green horizontal band) with center frequency $\omega_\mathrm{ro}$ and width that is 12.5\% of $\omega_\mathrm{ro}$ in this example. The green, red, and blue colors of the bands indicate intermodulation product classes with signal order $O_s=1$, 2, and 3, respectively, while higher order bands are gray and progressively more transparent. The width of each intermodulation product class is proportional to its signal order $O_s$. The pump tone, signals, idlers, and the dominant processes that can lead to readout crosstalk are labeled.
		}
	\label{fig:avoidance}
\end{figure}

Alternatively, many of the intermodulation products can be avoided by choosing the frequencies of the signals and the pump carefully.
To identify at which frequencies the intermodulation products occur, we divide the intermodulation products arising from amplification of $N$ frequency-multiplexed signals into classes with bounded bandwidth according to how many signal photons and which signs are involved in the process, see \cref{fig:avoidance}.
For example, if all signal frequencies $\omega_i$ are chosen between $\omega_\mathrm{min}$ and $\omega_\mathrm{max}$ (the signal band), and the pump is detuned by more than the width of the signal band
\begin{equation} \label{eq:pump-frequency-avoidance}
	\omega_p > 2\omega_\mathrm{max} - \omega_\mathrm{min},
\end{equation}
then none of the intermodulation frequency components of the form $\omega_i \pm (\omega_p - \omega_j)$, where $i, j \in {1...N}$, can collide with the signals.
This corresponds to the pump frequency in \cref{fig:avoidance} where the diagonal red bands do not overlap with the green horizontal signal band.
The main intermodulation products that can cause significant crosstalk fall into these two classes as they have signal order $O_s = 2$ and have therefore the highest amplitudes according to our measurements and \cref{eq:intermodulation-power-model}.
For typical widths of the signal band of about \SI{1}{GHz}, this implies signal-pump detunings of up to \SI{2}{GHz}. 
In \cref{fig:concept}~(c) we see that, while the gain of the TWPA is slightly reduced at large detunings from the pump, it is still sufficient to overcome the HEMT noise, as the signal-to-noise ratio remains constant over several~\si{GHz}.
We also see that overlap of the signal band with the $O_s=3$ band of intermodulation products of the class $\omega_i + \omega_j - \omega_k$ (blue horizontal band in \cref{fig:avoidance}) cannot be avoided and frequency collisions are likely for a high degree of multiplexing (see \cref{supp:collision-probability}).
For multiplexing the simultaneous readout of five qubits, see \cref{sec:implications}, it is feasible to choose readout frequencies to avoid all collisions with intermodulation products even if condition (\ref{eq:pump-frequency-avoidance}) is not satisfied.

In practice, a combination of frequency avoidance and suppression of intermodulation distortion will likely be required to achieve a high degree of multiplexing for fast readout with low crosstalk.

\section{Discussion and Outlook} \label{sec:discussion}

In this work we characterized the intermodulation distortion of a Josephson-junction-based traveling wave amplifier. 
We identified intermodulation products at frequencies of all integer prefactor combinations of two signal tone frequencies up to signal order $O_s = 3$ and characterized their dependence on the input power, which we found to be a power law with $O_s$ as the exponent if signal powers are swept together.

When operating the amplifier close to its saturation point with multiple input signals, we found that the signal-to-noise ratio is not much reduced compared to when using a single tone at the same total input power.
Intermodulation distortion can nonetheless lead to significant readout errors and crosstalk for multiplexed readout if an intermodulation product comes close to one of the signal frequencies.

Frequency collisions with intermodulation products of highest signal order $O_s = 2$ can be completely avoided if the pump frequency is designed to be detuned from the signals by more than the total width of the signal band.
The relative amplitude of higher-order intermodulation products can be suppressed by increasing the saturation power of the amplifier and by increasing the power-efficiency of readout.

A high degree of frequency multiplexing for readout is very desirable for scaling up general-purpose quantum processors to hundreds of qubits, e.g. for quantum error correction, which heavily relies on fast low-crosstalk mid-circuit measurements.
We found that careful consideration of the intermodulation distortion and amplifier saturation is needed to achieve the desired amplifier performance.

\section{Acknowledgments}

The authors thank Massachusetts Institute of Technology Lincoln Laboratory for providing the TWPA.
We acknowledge financial support by the Office of the Director of National Intelligence (ODNI), Intelligence Advanced Research Projects Activity (IARPA), through the U.S. Army Research Office grant W911NF-16-1-0071, by the EU Flagship on Quantum Technology H2020-FETFLAG-2018-03 project 820363 OpenSuperQ, by the National Center of Competence in Research `Quantum Science and Technology' (NCCR QSIT), a research instrument of the Swiss National Science Foundation (SNSF, grant number 51NF40-185902), by the SNSF R'Equip grant 206021-170731, by the EU programme H2020-FETOPEN project 828826 Quromorphic and by ETH Zurich. 
S.K. acknowledges financial support from Fondation Jean-Jacques et F\'{e}licia Lopez-Loreta and the ETH Zurich Foundation.
The views and conclusions contained herein are those of the authors and should not be interpreted as necessarily representing the official policies or endorsements, either expressed or implied, of the ODNI, IARPA, or the U.S. Government.

\begin{appendix}

\section{Experimental Setup}
\label{supp:setup}

\begin{figure}[t]
	\includegraphics[width=\columnwidth]{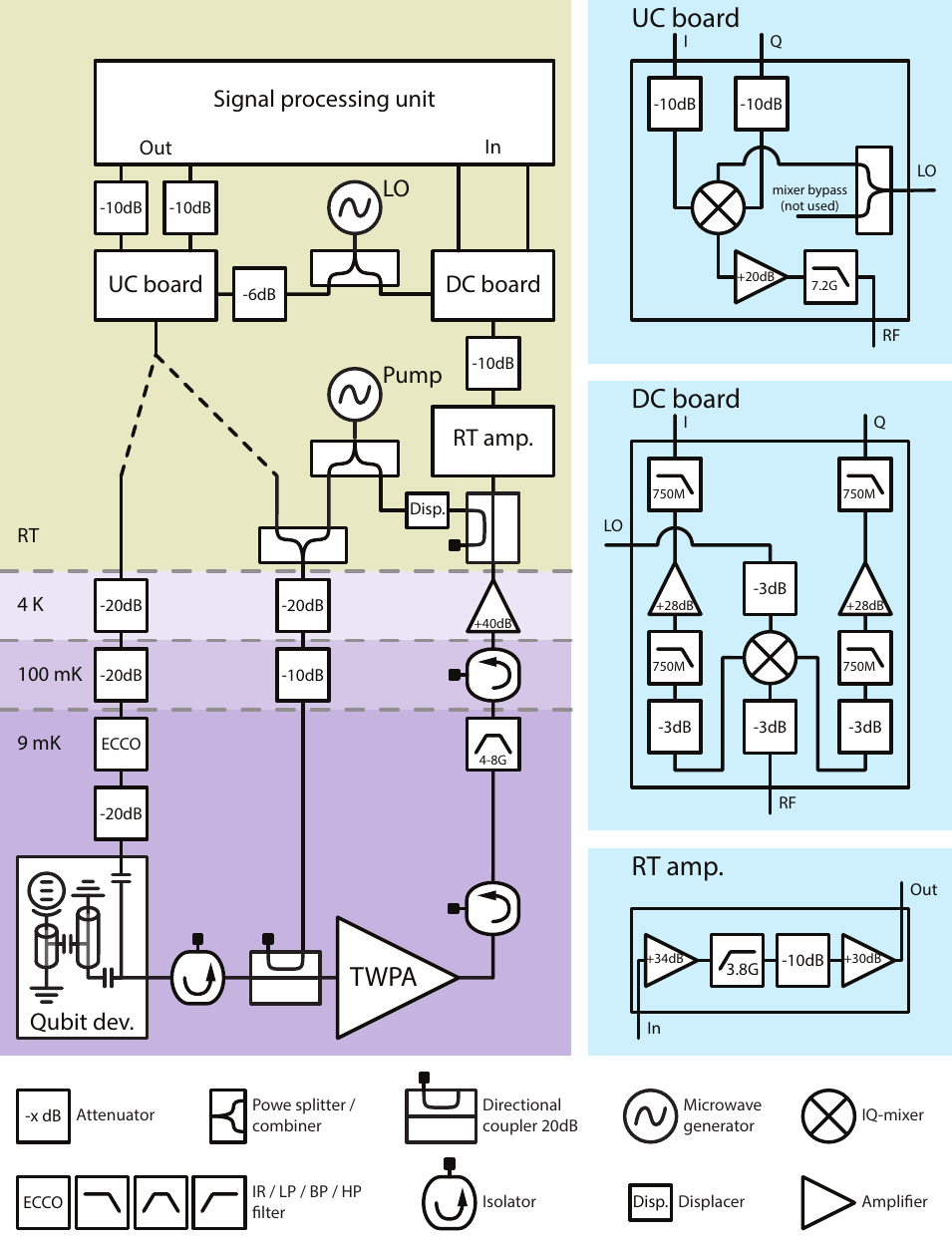}
	\caption{
		Schematic of the experimental setup. Probe signals from the up-conversion board are routed to the TWPA either via the directional coupler for TWPA characterization, or via the readout line of a quantum device with 17 qubits for qubit readout, as indicated by the black dashed connections. The temperature stage to which the components are thermalized is indicated by the different background colors.
		See text for details.
	}
	\label{fig:electronics-setup}
\end{figure}

We mount the TWPA and a quantum device with 17 transmon qubits at the base temperature stage of a dilution refrigerator, see \cref{fig:electronics-setup}.
The continuous-wave pump tone for the TWPA is directly generated using a microwave generator, while all the other signals are generated by a signal processing unit (UHF Quantum Analyzer) at an intermediate frequency between \SI{0}{MHz} and \SI{650}{MHz} and then up-converted (UC board) by IQ-modulation.
For qutrit readout we route the signals to a readout input port of the quantum device, where the readout resonators and their individual Purcell filters are coupled to five transmon qubits and a common feedline, see \cref{supp:qubit-device} for details.
The readout line output is amplified by the TWPA, which is mounted with a directional coupler at its input for injecting the pump tone, and sandwiched between a pair of dual junction wide-band isolators to remove back-propagating waves and to provide a \SI{50}{\ohm} matched environment.
For characterizing the TWPA performance separately, we bypass the quantum device by combining the signals with the pump tone and route all tones to the TWPA via the directional coupler.
The TWPA output is further amplified by a cryogenic HEMT amplifier and a pair of low-noise room-temperature amplifiers (RT amp.) before being down-converted (DC board) to intermediate frequency and digitized by the signal processing unit at \SI{1.8}{GS\per s}.

\section{Intermodulation Origin Verification}
\label{supp:origin-verification}

\begin{figure}[t]
	\includegraphics[width=\columnwidth]{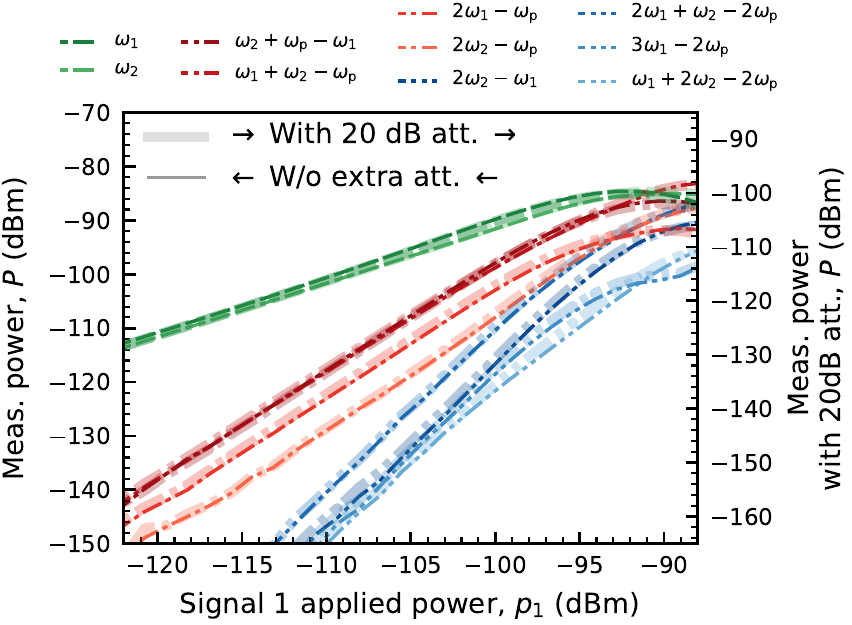}
	\caption{Comparison of measured powers $P$ with \SI{20}{dB} extra attenuation added before all room-temperature amplifiers. The powers with extra attenuation are shown on the right axis, which is shifted by \SI{15.1}{dB} relative to the left axis, see text.}
	\label{fig:power-sweep-with-attenuation}
\end{figure}

To verify that the intermodulation distortion which we observe originates from the TWPA, and not from any other element in the output line, we repeat the signal power sweep as shown in \cref{fig:characterization}~(c): first with the same configuration of the output line as in the rest of the paper, and once with an extra \SI{20}{dB} attenuator added before the first amplifier at room temperature.
In the second case, the input powers to all the room-temperature active elements are lowered, and the relative amplitudes of the intermodulation products compared to the signal levels would be lowered if the intermodulation distortion was due to any room-temperature component.
The cryogenic HEMT amplifier has a third order intercept power of $p_\mathrm{IP3,HEMT} = \SI{-40}{dBm}$, which is well above the intercept powers found for the TWPA.
We measure identical relative powers of the intermodulation products in the two datasets, see \cref{fig:power-sweep-with-attenuation}.
The absolute powers measured are \SI{15.1}{dB} reduced in the dataset with extra \SI{20}{dB} attenuation.
We attribute the \SI{4.9}{dB} difference from the expected \SI{20}{dB} shift to compression of the room-temperature low-noise amplifier by the strong pump tone, and correct for it when calculating the TWPA gain as a ratio of signal powers with the pump on and off.

\section{Additional Power Sweeps}
\label{supp:separate-powersweep}

\begin{figure}[t]
	\includegraphics[width=\columnwidth]{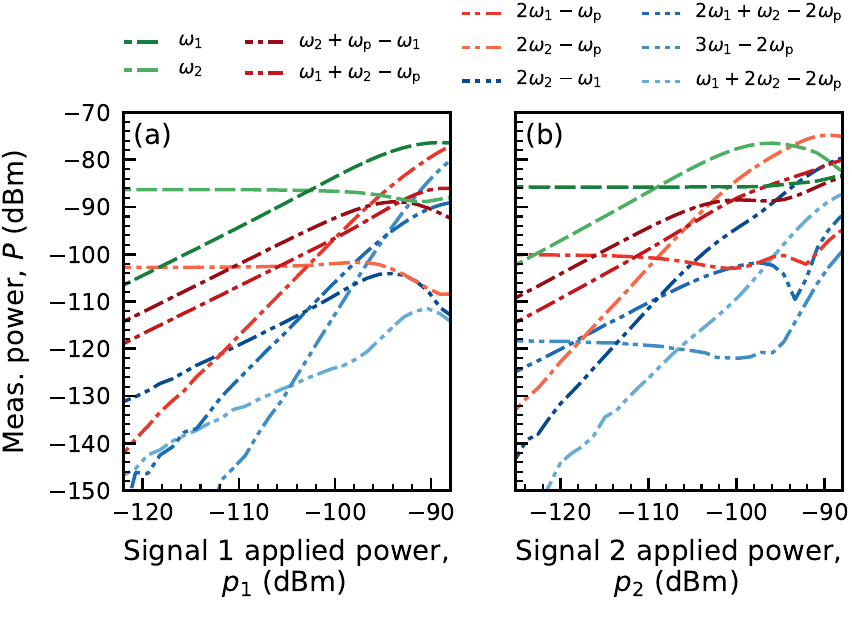}
	\caption{Power of the intermodulation products $P$ as a function of the applied signal power. In panel (a) the power $p_1$ of the tone at $\omega_1$ is swept while in panel (b) the power $p_2$ of the tone at $\omega_2$ is swept.}
	\label{fig:separate-powersweep}
\end{figure}

To verify the intermodulation power model~(\ref{eq:intermodulation-power-model}), we repeat the input power sweep experiment, keeping one of the input powers fixed: $p_2=\SI{-105}{dBm}$ or $p_1=\SI{-102}{dBm}$.
We find that below input powers that saturate the TWPA, the power in the intermodulation product at $n_1 \omega_1 + n_2 \omega_2 + n_p \omega_p$ follows a power law with $|n_1|$ or $|n_2|$ as the exponent when sweeping $p_1$ or $p_2$, respectively, see \cref{fig:separate-powersweep}. 
This is consistent with the model~(\ref{eq:intermodulation-power-model}).

\section{Single Tone Saturation}
\label{supp:saturation}

\begin{figure}[t]
	\includegraphics[width=\columnwidth]{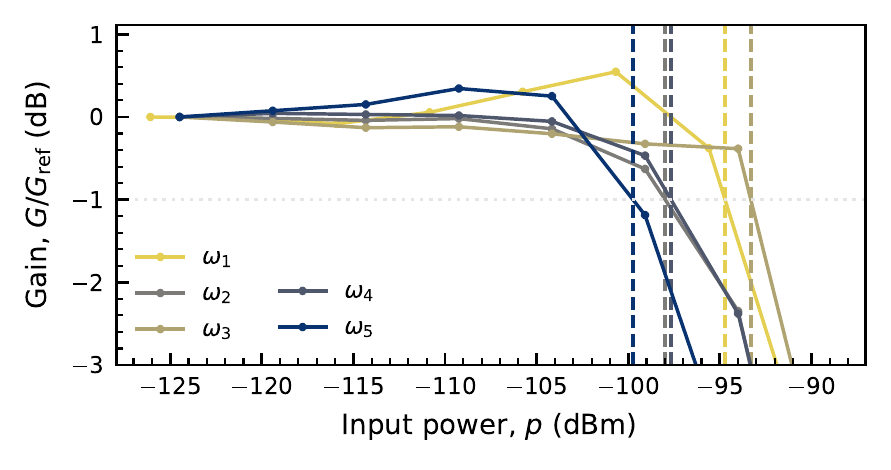}
	\caption{Change in gain $G$ compared to low-power reference values $G_\mathrm{ref}$ when applying a single tone at various frequencies as a function of the applied power $p$.
	The \SI{1}{dB} compression powers are indicated with vertical dashed lines.
	}
	\label{fig:saturation}
\end{figure}

To determine the mean \SI{1}{dB} compression point of the TWPA over the five signal frequencies $\{\omega_1,\ldots,\omega_5\}$, we apply each of the signals alone to the TWPA and vary their power, see \cref{fig:saturation}.
We find a mean \SI{1}{dB} compression power of $p_\mathrm{1dB} = \SI{-96.7\pm2.3}{dBm}$, with the uncertainty indicating the standard deviation over the different frequencies.
The average gain over the five signal frequencies here corresponds to the $N=1$ line in~\cref{fig:efficiency}~(a).

\section{Phase Dependence}
\label{supp:phase-dependence}

\begin{figure}[t]
	\includegraphics[width=\columnwidth]{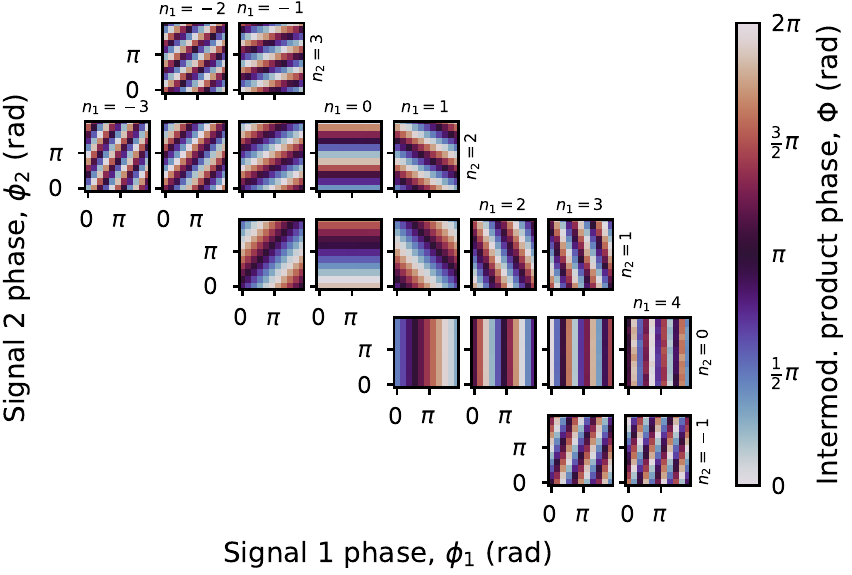}
	\caption{The phases of the intermodulation products at $n_1 \omega_1 + n_2 \omega_2 + (1 - n_1 - n_2) \omega_p$ when varying the phases $\phi_1$ and $\phi_2$ of the two input signals at $\omega_1$ and $\omega_2$.}
	\label{fig:phases}
\end{figure}

To determine whether the intermodulation products are phase coherent and how their phase depends on the phase of the input signals, we apply the two input tones at $\omega_1$ and $\omega_2$ at $p_1 = \SI{-102}{dBm}$ and $p_2 = \SI{-105}{dBm}$, respectively, and independently sweep their phases $\phi_1$ and $\phi_2$ between 0 and $2\pi$.
We extract the phase of the various intermodulation products from the output spectrum, see \cref{fig:phases}.
The phase $\Phi$ of each intermodulation product at $\omega = n_1 \omega_1 + n_2\omega_2 + n_p \omega_p$ follows the dependence
\begin{equation}
	\Phi = n_1 \phi_1 + n_2 \phi_2 + \theta,
\end{equation}
where $\theta$ is an intermodulation-product-specific phase offset.
For each $n_1$ and $n_2$, only the intermodulation product with $n_p = 1 - n_1 - n_2$ falls inside the acquisition bandwidth, and the intermodulation products not shown in \cref{fig:phases} are either outside the acquisition bandwidth or below the acquisition noise floor.


\section{Qubit Device}
\label{supp:qubit-device}

\begin{figure}[t]
	\includegraphics[width=\columnwidth]{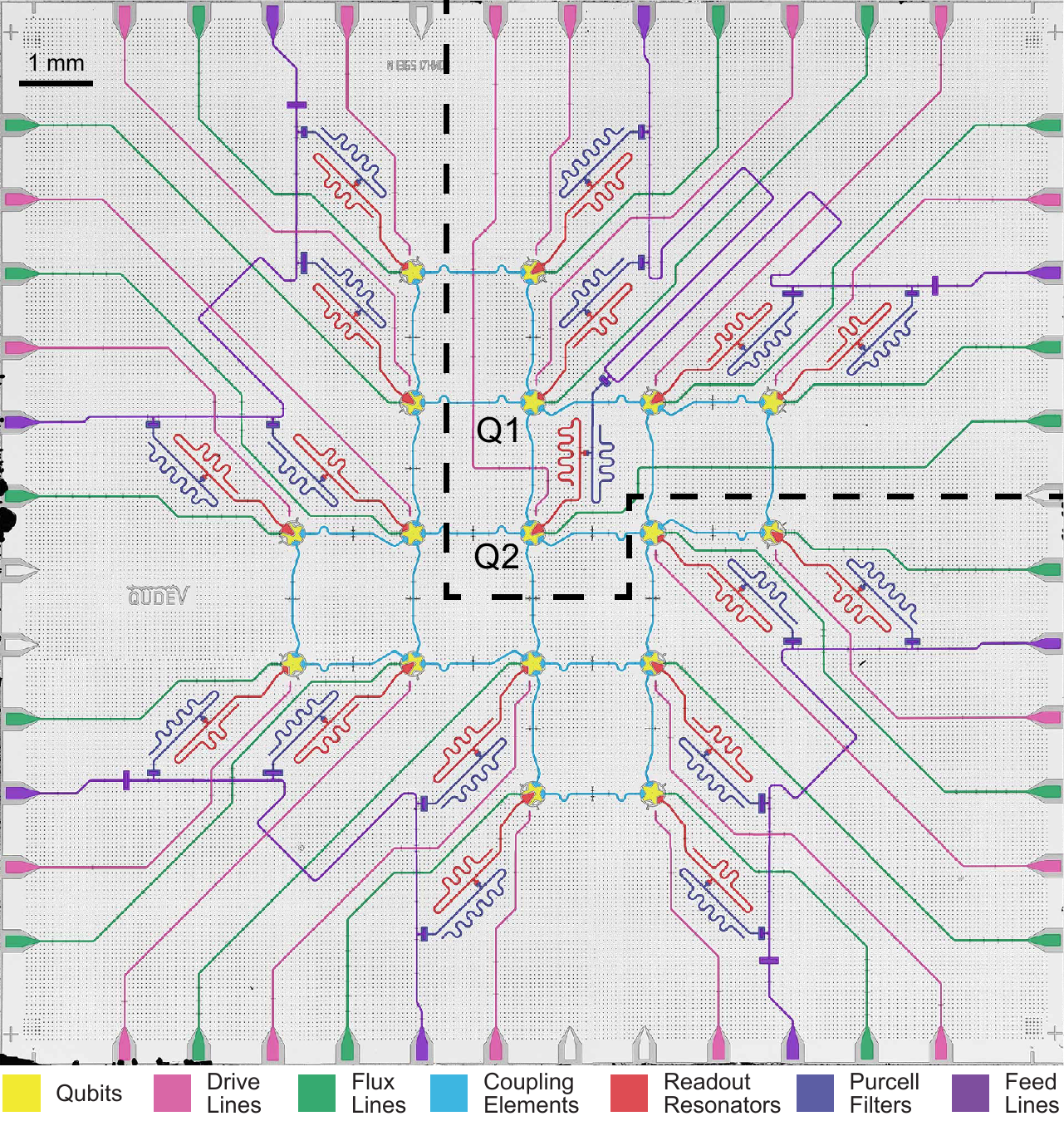}
	\caption{False-colored optical micrograph of the 17-qubit device used for the readout crosstalk investigation. For this work we used the feedline (purple) at the top-right corner of the device, separated by a dashed line, and two of the qubits coupled to it. Adapted from~\cite{Krinner2022}.}
	\label{fig:qubit-device}
\end{figure}

We use a subset of a 17-qubit device for characterizing qutrit readout in the presence of intermodulation distortion, see \cref{sec:implications} and the optical micrograph in \cref{fig:qubit-device}.
The transmon qubits have individual microwave (pink) and flux (green) control lines and readout resonators (red) with individual Purcell filters (blue) to suppress energy relaxation via the readout circuitry. 
The Purcell filters are coupled to a common readout feedline (purple), which is capacitively coupled on the input side and galvanically coupled on the output side to route the emitted readout signals to the output port.
The idling frequencies of Q1 and Q2 are \SI{5.88}{GHz} and \SI{3.90}{GHz}, lifetimes are \SI{17}{\micro s} and \SI{32}{\micro s}, and anharmonicities are \SI{-174}{MHz} and \SI{-186}{MHz}, respectively.
We bring Q2 to \SI{5.15}{GHz} for readout using a flux pulse to reduce the detuning from the readout resonator.

\section{Cross-Fidelity}
\label{supp:cross-fidelity}

The cross-fidelity as defined in the main text can be calculated as 
\begin{equation}
	F_{ij} = \frac{1}{d - 1}\left(\sum_{\xi \in \Xi_j} \max_{\zeta \in Z_i} \mathrm{Pr}(\xi | \zeta) - 1\right),
\end{equation}
where $d=3$ is the dimensionality of the qutrit, $\Xi_j = \{``g", ``e", ``f"\}$ are the classification outcomes of Q$j$, $Z_i = \{\ket{g}, \ket{e}, \ket{f}\}$ are the prepared states of Q$i$, and $\mathrm{Pr}$ is the probability distribution of classification results, conditioned on the prepared state of the qutrit.
This corresponds to the strategy of labeling the shot as the most likely state $\zeta$ of Q$i$, given the observed classification outcome $\xi$ of Q$j$

\section{Readout Circuit Power Efficiency}
\label{supp:power-efficiency}

To reduce the impact of intermodulation distortion, it is beneficial to reduce the signal power at the TWPA input.
One way to achieve this in the dispersive qubit readout scenario is to design  the readout circuit to maximize the power efficiency.
We quantify the power efficiency $\eta_p$ as the ratio of useful signal power to the maximum signal power, at which the useful signal power is calculated as the modulus squared of the difference of the output amplitudes for the $|g\rangle$ and $|e\rangle$ qubit states. 
Hence, we get the formula
\begin{equation}
	\eta_p = \frac{\left|\alpha^{(e)} - \alpha^{(g)}\right|^2}{4 \max\left(\left|\alpha^{(g)}\right|^2, \left|\alpha^{(e)}\right|^2\right)}.
\end{equation}
Here $\alpha^{(g)}$ and $\alpha^{(e)}$ are the steady state amplitudes of the signals input to the TWPA when the qubit is in the $|g\rangle$ or $|e\rangle$ state, respectively.
The normalization is chosen such that $\eta_p = 1$ in the optimal case $\alpha^{(e)} =- \alpha^{(g)}$.

Our readout circuit consists of $\lambda/4$ readout resonators coupled to a feedline via individual $\lambda/4$ Purcell filters, with coupling $J$ between the resonators and a linewidth $\kappa_P$ of the Purcell filter~\cite{Heinsoo2018}.
Because of the strong coupling between the two resonators $J \approx \kappa_P/2$, we use just one of the two eigenmodes of the two-resonator system, which behaves approximately like a single resonator side-coupled to the feedline.
The power efficiency for such a system is given by $\eta_{p} = \kappa^2 / (\kappa^2 + 4 \chi^2)$, where $\kappa$ and $\chi$ are the effective linewidth and the qubit-state-dependent dispersive shift of the eigenmode.
For this system, the power efficiency improves as $\chi$ decreases.
On the other hand, for $J \ll \kappa_P$, the readout resonator mode behaves like a single $\lambda/2$ resonator measured in transmission, in which case the power efficiency $\eta_{p} = 4 \chi^2 / (\kappa^2 + 4 \chi^2)$ increases with increasing $\chi$.
For both cases $\eta_{p}=0.5$ is achieved for $\chi = \kappa/2$, which maximizes the steady state readout speed for a fixed number of photons in the readout mode~\cite{Walter2017}.
We thus find that the power efficiency can strongly depend on the circuit parameters and each case needs to be analyzed in detail.
Independent of the circuit parameters, the power efficiency can be brought to 100\% by displacing the signal with the negative of the mean response over the two states: $\alpha^{(g)} \rightarrow \alpha^{(g)} - (\alpha^{(g)} + \alpha^{(e)})/2 = (\alpha^{(g)} - \alpha^{(e)})/2$, and $\alpha^{(e)} \rightarrow (\alpha^{(e)} - \alpha^{(g)})/2$.

\section{Collision Probability}
\label{supp:collision-probability}

\begin{figure}[t]
	\includegraphics[width=\columnwidth]{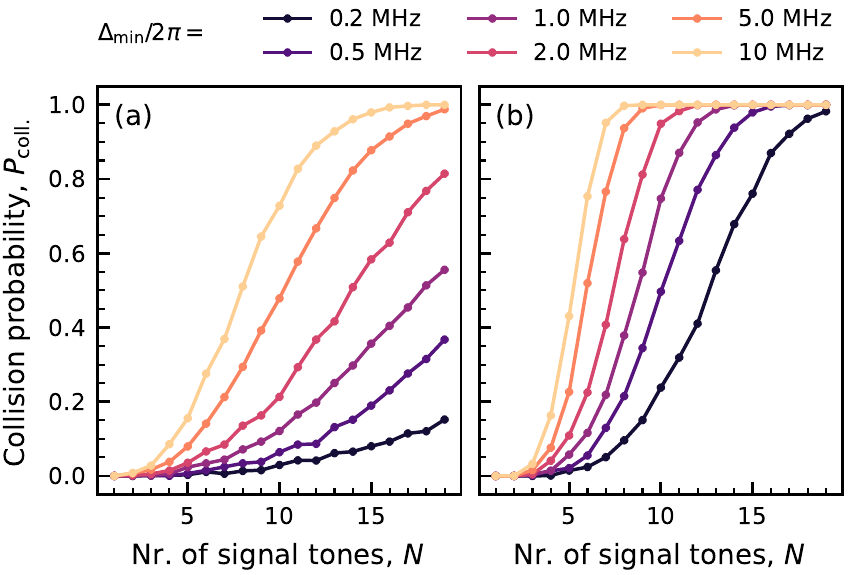}
	\caption{
		(a) Probability $P_\mathrm{coll.}$ of an intermodulation product of signal order $O_s = 2$ colliding with a readout tone for different minimal allowed detunings $\Delta_\mathrm{min}$ and degrees of multiplexing $N$. The readout frequencies are sampled from a uniform distribution within a \SI{1}{GHz} readout band with a minimum detuning of \SI{20}{MHz}. The pump is detuned by \SI{0.52}{GHz} from the edge of the signal band.
		(b) Same as (a), but for $O_s=3$ intermodulation products and a pump detuning of \SI{2.05}{GHz}.
	}
	\label{fig:collision-probability}
\end{figure}

To assess the difficulty of avoiding frequency collisions with intermodulation products, we determine the probability of a collision for various degrees of readout multiplexing $N$.
We sample 2000 signal frequency configurations from a uniform distribution between \SI{6.4}{GHz} and \SI{7.4}{GHz} with a minimal detuning of \SI{20}{MHz} between any signals.
If any intermodulation product has a smaller detuning from a signal than $\Delta_\mathrm{min}/2\pi = \{0.2, 0.5, 1.0, 2.0, 5.0, 10\} \si{MHz}$, we consider it as a collision. 
These detunings correspond to the full-width half-maximum bandwidths of square pulses of lengths $\{3, 1.2, 0.6, 0.3, 0.12, 0.06\} \si{\micro s}$.
First, we consider a pump frequency of $\omega_p/2\pi = \SI{7.92}{GHz}$, in which case condition~(\ref{eq:pump-frequency-avoidance}) is not satisfied, and we calculate the probability of collision with $O_s=2$ intermodulation products, see \cref{fig:collision-probability}~(a).
Second, we consider a pump frequency of $\omega_p/2\pi = \SI{9.45}{GHz}$, for which there are no collisions with $O_s=2$ intermodulation products and we calculate the collision probability with $O_s=3$ terms, see~\cref{fig:collision-probability}~(b).

By carefully choosing the readout frequencies, it is possible to avoid all $O_s=2$ intermodulation products when multiplexing 120-ns-long readout of up to $N=10$ qubits even with~(\ref{eq:pump-frequency-avoidance}) not satisfied, as the collision probability is around 50\% for random sampling of readout frequencies (see $\Delta_\mathrm{min}/2\pi=\SI{5}{MHz}$ line in \cref{fig:collision-probability}~(a))
However, if the readout frequencies differ from the design values on the order of \SI{10}{MHz}, e.g., because of variations in the resonator fabrication, avoiding collisions becomes almost impossible (see $\Delta_\mathrm{min}/2\pi=\SI{10}{MHz}$ line in \cref{fig:collision-probability}~(a)).
On the other hand, avoiding $O_s=3$ intermodulation products becomes practically impossible for multiplexing 120-ns readout for more than $N=6$ qubits.
This means that for fast high-fidelity readout with a large degree of multiplexing, it is critical to make sure that the third order intercept power of the amplifier $p_\mathrm{IP3}$ is well above the used signal power.

To be more specific, let us consider the collision probability for 120-ns-long readout pulses when the condition (\ref{eq:pump-frequency-avoidance}) is not satisfied for different sizes of surface codes~\cite{Fowler2012,Tomita2014}. 
For a distance-3 device with 17 qubits, the total failure probability is 18\% when using four multiplexed readout lines, as was done for the quantum device here.
For a distance-5 device with 49 qubits with similar degree of multiplexing (12 multiplexed readout lines), the failure probability increases to 40\%, and quickly increases if the degree of multiplexing is increased: 54\% for 10 readout lines and 73\% for 8 readout lines.

\end{appendix}

\bibliography{references}


\end{document}